\documentstyle [preprint,prb,aps,eqsecnum]{revtex}

\newcommand{\be}{\begin{equation}}
\newcommand{\ee}{\end{equation}}

\begin {document}
\draft
\title{Critical and bicritical properties of Harper's equation with next
nearest neighbor coupling}

\author {J. H. Han and D. J. Thouless}
\address{ Department of Physics, FM-15,
    University of Washington, Seattle, Washington 98195 }

\author {H. Hiramoto}
\address{College of Humanities and Sciences, Nihon University,
Sakurajoshui, Setagaya-ku, Tokyo 156, Japan}

\author{M. Kohmoto}
\address{Institute for Solid State Physics, University of Tokyo, 7-22-1
Roppongi, Minato-ku, Tokyo 106, Japan}
\date{\today}

\maketitle

\begin {abstract}
We have exploited a variety of techniques to study the universality and
stability of the scaling properties of Harper's equation, the
equation for a particle moving on a tight-binding square lattice in the
presence of a gauge field, when coupling to next nearest sites is added.
We find, from numerical and analytical studies, that the scaling
behavior of the total width of the spectrum and the multifractal nature of
the spectrum are unchanged, provided the next nearest neighbor coupling terms
are below a certain threshold value. The full square symmetry of the
Hamiltonian is not required for criticality, but the square diagonals should
remain as reflection lines. A bicritical line is found at the boundary
between the region in which the nearest neighbor terms dominate and the
region in which the next nearest neighbor terms dominate. On the bicritical
line a different critical exponent for the width of the spectrum and
different multifractal behavior are found. In the region in which the next
nearest neighbor terms dominate the behavior is still critical if the
Hamiltonian is invariant under reflection in the directions parallel to the
sides of the square, but a new length scale enters, and the behavior is no
longer universal but shows strongly oscillatory behavior. For a flux per unit
cell equal to $1/q$ the measure of the spectrum is proportional to $1/q$ in
this region, but if it is a ratio of Fibonacci numbers the measure decreases
with a rather higher inverse power of the denominator.
\end {abstract}
\narrowtext

\newpage
\tighten
\baselineskip = 24pt

\section{Introduction}

Harper's equation can be derived as the equation for an electron in a strong
two-\newline
dimensional periodic potential and a weak magnetic field, or for an electron
in a strong magnetic field and a weak periodic potential.
The Hamiltonian can be written in the form
\be
 H(p_x,x)=2t_a\cos p_x+2t_b\cos x\;,
\label{eq:harp} \ee
where
\be
 p_x=-2\pi i\phi d/dx
\label{eq:pxdef}\ee
is the variable conjugate to $x$. Azbel\cite{azbel64} showed that in the case
$t_a=t_b$,
which corresponds to a periodic potential with square symmetry, the spectrum
forms a ``devil's staircase'' for irrational values of $\phi$, and Hofstadter
\cite{hofstadter76} generated computer drawings of the spectrum for rational
values of $\phi$,
and discussed the self-similarity and scaling of the spectrum. Work by Aubry
and Andr\'e\cite{aubry80} exploited the symmetry of the problem under canonical
transformations that interchange $x$ and $p_x$, and showed that, for irrational
values of $\phi$, there is a localization length in the $x$ direction,
independent of energy,
\be
 L=1/\ln(t_b/t_a),\ \  {\rm for}\ \ t_b>t_a,
\ee
which diverges at this symmetry point $t_a=t_b$. Also the sum of the widths of
all the energy bands, the measure of the spectrum, has the form
\be
W=4|t_b-t_a|\;,
\label{eq:width}\ee
which vanishes at the same point. These properties have suggested that this
point is like a critical point of the system, that $|1-t_a/t_b|$ represents the
distance from the critical point, and that, where $\phi=p/q$ is a rational,
the denominator $q$ acts as a finite size in finite size scaling theory
\cite{thouless83}.

Various methods have been used to study the Harper equation. There are some
rigorous analytical results on the Lyapunov exponent (reciprocal of the
localization length)\cite{souillard87}, and on the measure of the spectrum (sum
of
the band widths)\cite{avron90,thouless91a,watson91,last92,last93}.
Aubry duality gives information about the localization length.

If $H(p_x,x)$ defined in Eq.\ (\ref{eq:harp})
is treated as a classical Hamiltonian there
is an obvious interpretation of the critical point, since for $t_a>t_b$ the
energy contours surrounding the minima of $H$ are separated from the contours
surrounding the maxima of $H$ by a region of orbits open in the $x$ direction,
while for $t_a<t_b$ there are orbits open in the $p_x$ direction. For the
symmetric case $t_a=t_b$ there is a single energy $E=0$ separating the two
types of closed orbits. A more subtle semiclassical analysis, based on scaling
theory, must be introduced to explain why the behavior is not just singular at
the energy of this separatrix, but is singular at all energies in the spectrum.

Much of the information we have about this problem comes from numerical studies
of the spectrum for rational values of $\phi$ similar to the one carried out
by Hofstadter. The vanishing of the measure of the spectrum at the critical
point gives a one-parameter indicator of the critical point. Numerical studies
on the Harper equation indicate that this measure is given by Eq.\
(\ref{eq:width}) for all
irrational values of $\phi$, and that for rational values $\phi=p/q$ the
measure scales as\cite{thouless83}
\be
 W\asymp (t_b-t_a)g[q(1-t_a/t_b)]\;,
\label{eq:scale} \ee
where the scaling function $g(s)$ behaves as $9.3299/s$ when its argument
is small, and tends to $\pm 4$ for large $|s|$. Only the corrections to
scaling seem to be sensitive to the value of $p$\cite{thouless91a,thouless91b}.
The distribution of band
widths within this spectrum gives a more detailed picture of the spectrum, but
this is sensitive to the value of $\phi$. The simplest forms might be
expected if $\phi$ is an irrational solution of a quadratic equation, so
that its continued fraction expansion repeats itself, or if it is a rational
approximant obtained by truncating such a repeating continued fraction. The
golden mean, which is the limit of the ratios of successive Fibonacci numbers,
is the simplest case of this sort. In this case the spectrum has a
multifractal structure which is self-similar at the critical point
\cite{hofstadter76,kohmoto83,tang86,bell89}.

Another method which has been used to get a number of important results takes
its simplest form when $\phi$ is small, so that Eq.\ (\ref{eq:harp}) can be
treated semiclassically.
 It is then a second order difference equation with a slowly
varying central term, for which the WKB equation can be used. For
$t_b\approx t_a$ only those bands close to $E=0$ have appreciable band width,
and the others all have widths which vanish exponentially as $1/\phi$ gets
large, since the widths are determined by tunnelling between successive minima
or maxima of $2t_b\cos x$. The case $\phi=1/q$ is particularly
simple, and this has been used to derive an analytic form for the function
$g(s)$ of Eq.\ (\ref{eq:scale})\cite{thouless90,thouless91a,watson91}.

The critical properties of the Harper equation have been explored in great
detail by various combinations of these methods. However, it is not clear
to what extent the elegant properties of the Harper equation are special
properties of that equation, and to what extent they are robust, and
survive perturbations of the model. On the basis of the small $\phi$
behavior, Suslov\cite{suslov82} has argued that modifications of the
$x$-dependence of
Eq.\ (\ref{eq:harp}) which maintain the periodicity will lead to an
energy-dependence of the critical value of
$t_b$.
This behavior is supported by numerical
calculations.\cite{soukoulis82,hiramoto89a,hiramoto89b}
On the other hand Helffer
and Sj\"ostrand\cite{helffer91} have argued  that variants of Eq.\
(\ref{eq:harp}) which
preserve the invariance under the canonical transformation $p_x\to -x$,
$x\to p_x$ are critical at all energies. It is the purpose of this work to
take these questions further by exploring in some detail a simple
generalization of the Harper equation.

The equation we explore is the generalization of Eq.\
(\ref{eq:harp}) which takes the form
\be
 H(p_x,x)=2t_a\cos p_x+2t_b\cos x +2t_{a\bar b}\cos(p_x-x)
+2t_{ab}\cos(p_x+x)\;,
\label{eq:harpnnn}\ee
This could represent a tight-binding model in which electrons can tunnel to
next nearest neighbors as well as to nearest neighbors on a rectangular
lattice, and, for
$t_{ab}=0$, $t_a=t_b=t_{a\bar b}$, it can represent an electron on a
triangular lattice\cite{claro79}. This form of the Hamiltonian is very
convenient for numerical work, as it remains in the form of a three-term
difference equation,and there have been a number of earlier studies of it
\cite{thouless83,claro79,hatsugai90}. It has enough parameters
that we can study the effects of different features of the Hamiltonian such as
the breaking of the symmetry, and the change in the nature of the constant
energy contours. For $t_a=t_b$ and $t_{a\bar b}=t_{ab}$ the system has
square symmetry, but there are two quite different regimes with this
symmetry, according to whether $t_a$ or $t_{ab}$ is dominant. We show
that there is actually an interesting bicritical point which separates the
two different regimes.

In this paper we have exploited
most of the techniques mentioned above. In Sec.\ II we discuss the
main features of the energy contours of Eq.\ (\ref{eq:harpnnn}) and discuss
the symmetries of the system. In Sec.\ III we discuss the form of the
characteristic equation whose roots give the eigenvalues, repeat some earlier
results which were
obtained by the use of Aubry duality, discuss the extension of finite size
scaling arguments to this case, and extend the sum rule of Last and
Wilkinson\cite{last92} to this case. In Sec.\ IV we present an analysis
in terms of multifractals for the case that $\phi$ is a ratio of
neighboring Fibonacci numbers. In Sec.\ V we give the results of
calculations of the measure of the spectrum when $\phi$ is a fraction such
as $1/q$ or $p/(p^2+1)$ that represents a slow modulation of the diagonal
term of the difference equation. These two approaches complement one
another, since for the ratio of Fibonacci numbers the spectrum is spread
over a large number of bands, even close to the critical point, whereas for
small values of $\phi$ the only bands with significant measure are close to
zero energy. We find that the behavior is relatively simple in the region
in which $t_a=t_b$ is dominant, but there is a bicritical region not only
at $t_a=t_b=2t_{a\bar b}=2t_{ab}$, but also for $t_a=2t_{a\bar
b}=2t_{ab}>t_b$. For $t_{a\bar b}=t_{ab}$ dominant the situation appears to
be much more complicated, with some important oscillatory terms which
confuse the analysis of numerical results.

In Sec.\ VI we do what we can to explain the results we have from numerical
analysis in terms of WKB theory and scaling theory. In some cases our
understanding is reasonably complete, but in other cases we can do little more
than explain why the problem is complicated. There is a concluding
discussion in Sec.\ VII.

\section{Classical orbits and symmetry}

In this discussion we assume all the coefficients $t_a$, $t_b$, $t_{a\bar
b}$, $t_{ab}$ are positive. For the case $t_a=t_b$, $t_{a\bar b}=t_{ab}$
the spectrum of the Hamiltonian given by Eq.\ (\ref{eq:harpnnn}) is
invariant under the eight operations of the symmetry group of the square.
The four proper rotations are generated by $p_x\to -x$, $x\to p_x$, while
the time reversal operation $p_x\to -p_x$ generates the improper rotations.
The Hamiltonian is also invariant under the group of translations in phase
space corresponding to a square lattice. The classical Hamiltonian has a
maximum at $p_x=0=x$, where its value is $4(t_a+t_{ab})$, and at
equivalent lattice points. For $t_a>2t_{ab}$ it has a minimum at
$p_x=\pi=x$, where its value is $-4(t_a-t_{ab})$. There are two saddle
points in each unit cell, at $p_x=0$, $x=\pi$, and at $p_x=\pi$, $x=0$,
where its value is $-4t_{ab}$. At this value of the energy there is a
contour given by
\be
\cos p_x=-{2t_{ab}+t_a\cos x\over t_a+2t_{ab}\cos x}
\label{eq:cospx}\ee
which threads the system, separating contours that surround minima from those
that surround maxima. For $t_a<2t_{ab}$ the maximum is unchanged, but
the points at $p_x=\pi=x$ become subsidiary maxima, while the points at
$p_x=0$, $x=\pi$, and at $p_x=\pi$, $x=0$ become minima. Four more saddle
points appear at $\cos p_x=-t_a/2t_{ab}$, $\cos x=-t_a/2t_{ab}$,
where the energy is $-t_a^2/t_{ab}$. The contour joining the saddle
points and separating contours
surrounding minima from those surrounding maxima is now
\be
 \cos p_x=-t_a/2t_{ab}\ \ {\rm or}\ \ \cos x= -t_a/2t_{ab}\;.
\label{eq:cospx2}\ee
In this paper we pay particular attention to the bicritical point
$t_a=t_b=2t_{a\bar b}=2t_{ab}$, where all contours surround maxima except
for the lines where the energy has its minimum value $-2t_a$. At the points
$p_x=\pi=x$ the lowest nonvanishing partial derivatives of the energy are
the fourth
derivatives.

For the case $t_a=t_b$, $t_{a\bar b}>t_{ab}$ the symmetry operation
$p_x\leftrightarrow x$, a reflection symmetry in phase space, remains, as
well as rotation by $\pi$. There is still a maximum at $p_x=0=x$, where the
energy is $4t_a+2t_{a\bar b}+2t_{ab}$, and at equivalent lattice points.
For $t_a>t_{a\bar b}$ there is a minimum at $p_x=\pi=x$, where its value is
$-4t_a+2t_{a\bar b}+2t_{ab}$. For $t_a^2>4t_{a\bar b}t_{ab}$ there are two
saddle points in each unit cell, at $p_x=0$, $x=\pi$ and at $p_x=\pi$,
$x=0$, where its value is $-2(t_{a\bar b}+t_{ab})$. At this value of the
energy there is a contour given by
\be
 2t_{ab}\cos [{1\over 2}(p_x+x)]=-\cos [{1\over
2}(p_x-x)][t_a\pm\sqrt{t_a^2-4t_{a\bar b}t_{ab}}]
\label{eq:contour}\ee
which threads the system, provided $t_a>t_{a\bar b}+t_{ab}$. For
$t_a<t_{a\bar b}+t_{ab}$ there is a range of energies in the neighborhood
of $-2(t_{a\bar b}+t_{ab})$, with $\cos [{1\over 2}(p_x-x)]$ near $\pm 1$,
for which there are no values of $(p_x+x)$ that satisfy Eq.\
(\ref{eq:contour}). In this case there are open orbits in the direction of
constant $(p_x- x)$.

A special case of this sort is $t_a=t_b=t_{a\bar b}$, $t_{ab}=0$, which has
triangular symmetry, and is equivalent to the case worked out numerically
by Claro and Wannier\cite{claro79}.

For $t_a\ne t_b$, $t_{a\bar b}=t_{ab}$ the Hamiltonian is invariant under
$p_x\to-p_x$. For $t_{a\bar b}=t_{ab}>t_b>t_a$ there are maxima at
$p_x=0=x$ and at $p_x=\pi=x$, where the energy has the values $4t_{ab}
\pm(t_a+t_b)$, and minima at $p_x=0$, $x=\pi$ and at $p_x=\pi$, $x=0$,
where the energies are $-4t_{ab}\pm(t_a-t_b)$. There are saddle points
given by
\be
 \cos p_x=-t_b/2t_{ab},\ \ \cos x=-t_a/2t_{ab}\;,
\label{eq:contour2}\ee
and the energy has the value $-t_at_b/t_{ab}$ on the lines on which
either of these conditions is satisfied. For $t_b>2t_{a\bar b}=2t_{ab}>t_a$
the four saddle points given by Eq.\ (\ref{eq:contour2}) disappear, and the
new saddle points are at $p_x=\pi$, $x=0$ and  at $p_x=\pi=x$, where the
energies are $-t_a\pm(2t_{ab}-t_b)$. There are orbits open in the
$p_x$ direction between these energies.

\section{\bf Characteristic equation and Aubry duality}

{}From Eq.\ (\ref{eq:pxdef}) it can be seen that the operator $2\cos p_x$ is
a displacement operator that displaces the coordinate by $2\pi\phi$. The
eigenvalue problem for the Hamiltonian (\ref{eq:harpnnn}) takes the form of
a set of finite difference problems
\[
(t_a+t_{a\bar b}e^{{2\pi i\phi}(n-{1\over 2})+ik_2} +t_{ab}e^{-{2\pi
i\phi}(n-{1\over 2})-ik_2}) a_{n-1} +2t_b\cos({2\pi \phi n}+k_2)a_n
\]
\be
+(t_a+t_{a\bar b}e^{-{2\pi i\phi}(n+{1\over 2})-ik_2} +t_{ab}e^{{2\pi
\phi}(n+{1\over 2})+ik_2})a_{n+1} =Ea_n\;,
\label{eq:nnn}\ee
with the variable parameter $k_2$ determined by the initial value of $x$. When
$\phi=p/q$ is rational this equation is periodic with period $q$, and
solutions of the Floquet form
\be
a_n =c_ne^{ik_1n} \;,
\label{eq:ft}\ee
with $c_n$ periodic, can be found. This then yields a finite matrix problem,
with the matrix tridiagonal apart from the top right and bottom left corners,
for which the characteristic polynomial has as its only $k_1$ dependent term
\be
 (-1)^{q-1}e^{ik_1q}\prod_{n=0}^{q-1}\bigl(t_a+t_{a\bar b}e^{-{2\pi ip\over q}
(n+{1\over 2})-ik_2} +t_{ab}e^{{2\pi ip\over q}(n+{1\over 2})+ik_2} \bigr)+{\rm
conjugate\ complex}\;.
\label{eq:offd} \ee
This product can  be expressed in terms of a Chebyshev polynomial $T_q$ (see
Appendix A), and this gives
\[
(-1)^{q-1}4t^q \cos(qk_1) T_q({t_a\over
2t})
\]
\be
  +(-1)^p2\{t_{a\bar b}^q\cos[q(k_1-k_2)] +t_{ab}^q\cos[q(k_1+k_2)]\} \;,
\ee
where
\be
t^2=t_{a\bar b}t_{ab}\;.
\ee
Since the whole spectrum is, from Eq.\ (\ref{eq:harpnnn}), clearly
invariant under the interchange of $p_x$, $x$ and $t_a$, $t_b$, there must
also be a similar $k_2$ dependent term in the characteristic polynomial,
so the characteristic polynomial can be written in the form
\[
P(E)=P_0(E)-(-1)^q4t^q\{\cos(qk_1)
T_q({t_a\over 2t})+\cos(qk_2) T_q({t_b\over
2t})\}
\]
\be
+(-1)^p2\{t_{a\bar b}^q\cos[q(k_1-k_2)] +t_{ab}^q\cos[q(k_1+k_2)]\} \;,
\label{eq:offdiag} \ee
where $P_0(E)$ is independent of $k_1$, $k_2$. The energy bands are
determined by the variation of the solutions of the characteristic equation
as $k_1$, $k_2$ are varied.

Some simple analysis can show which of the terms in this expression will
dominate in the limit of large $q$. For $x>1$ we have
\be
 [T_q(x)]^{1/q}\approx x+\sqrt{x^2-1}\;,
\ee
and so, for $t_b\ge t_a$, $t_{a\bar b}>t_{ab}$, the dominant term in
Eq.\ (\ref{eq:offdiag}) is of order
\be
 [{1\over 2}(t_b+\sqrt{t_b^2-4t^2})]^q
\ee
for $t_b>t_{a\bar b}+t_{ab}$, and is of order $t_{a\bar b}^q$ for $t_b<t_{a\bar
b}+t_{ab}$. This transition from a regime where the band widths are dominated
by
the term depending on $k_2$ to a regime where the band widths are dominated by
a
term depending on $k_1-k_2$ occurs at the same values of the parameters as the
changes in the nature of the classical orbits which we discussed in connection
with Eqs. (\ref{eq:contour}) and (\ref{eq:contour2}).
Under these conditions only one term in Eq.\ (\ref{eq:offdiag}) is
relevant for large $q$, or two terms if $t_a=t_b$. Since this expression is
independent of the value of the energy, one should expect the critical values
of
the parameters to be energy independent, whereas the classical orbits only give
information about the behavior at the singular value of the energy.

For $2t_{a\bar b}=2t_{ab}>t_b\ge t_a$ the situation is very different,
since the Chebyshev polynomials are now of order unity, and all four terms
in Eq.\ (\ref{eq:offdiag}) would seem to be marginal for large $q$. In
particular one should expect the band widths to depend on
\be
\cos[q\arccos({t_a/ 2t_{ab}})] \ \ {\rm and} \ \ \cos[q\arccos({t_b/
2t_{ab}})]\;,
\label{eq:newlength} \ee
so the band widths should display nearly periodic behavior in $q$. In fact we
found such behavior in the numerical studies reported in this paper before we
had realised that they ought to be found.

In an earlier work\cite{thouless83} it was shown how the argument of Aubry
and  Andr\'e can be adapted to this situation. There are three parts to this
argument. Firstly they state that the element of the Green function
connecting the two ends of a tridiagonal matrix can be expressed as the
product of the next-to-diagonal matrix elements divided by the characteristic
polynomial. The product of off-diagonal matrix elements is just the
coefficient of $\cos p_x$ in Eq.\ (\ref{eq:offdiag}). Secondly they compare
this expression with the expression for the dual problem obtained by
interchanging $p_x$ and $x$. The characteristic polynomial is unchanged,
and the ratio of the products of next-to-diagonal elements can be used to
generate an expression for the difference between the Lyapunov exponents in
the $x$ and $p_x$ directions, in the form
\be
 \lambda_x-\lambda_p= \lim_{q\to\infty} {1\over q}\ln\bigl[
{4t^q |T_q({t_b\over 2t})|
+2t_{a\bar b}^q+ 2t_{ab}^q \over 4t^q |T_q({t_a\over
2t})| +2t_{a\bar b}^q +2t_{ab}^q }\bigr]\;.
\label{eq:lyap} \ee
This is zero for $t_{a\bar b}+t_{ab}\ge t_b\ge t_a$. For $t_b\ge t_a\ge
t_{a\bar b}+t_{ab}$ it gives
\be
 \lambda_x-\lambda_p= \ln\bigl[{{1\over 2}t_b+{1\over
2}\sqrt {t_b^2-4t^2}
 \over {1\over 2}t_a+{1\over 2}\sqrt {t_a^2-4t^2}}\bigr] \;,
\label{eq:lyap1} \ee
and for $t_b\ge t_{a\bar b}+t_{ab}\ge t_a$, $t_{a\bar b}\ge t_{ab}$ it
gives
\be
 \lambda_x-\lambda_p= \ln\bigl[{{1\over 2}t_b+{1\over 2}\sqrt
{t_b^2-4t^2} \over t_{a\bar b}}\bigr] \;.
\label{eq:lyap2} \ee
The third part of the argument, which we find to be rather more subtle,
says that if $\lambda_x$ is positive then $\lambda_p$ must be zero, so
Eqs.\ (\ref{eq:lyap1}) and (\ref{eq:lyap2}) are actually equations for
$\lambda_x$ rather than for $\lambda_x-\lambda_p$. Eigenstates for the
Hamiltonian (\ref{eq:harpnnn}) in $p_x$ space can be found from
eigenstates in $x$ space by Fourier transformation. These eigenstates in
$x$ space have their support on a lattice of points, so their Fourier
transforms are periodic. If they are localized in space their Fourier
transforms are smooth periodic functions. Functions of this sort
corresponding to the same value of the energy cannot be superposed to give
localization in $p_x$ space as well as localization in $x$ space.

These results show that states are localized in $x$ space, independent of
energy, for generic irrational $\phi$ provided $t_b$ is greater than both
$t_a$ and $t_{a\bar b}+t_{ab}$. This condition, now independent of energy,
is the same as the condition for the existence of open orbits extended in
the $p_x$ direction given in the discussion of Eq.\ (\ref{eq:contour2}).

The finite size scaling argument that lead to Eq.~(\ref{eq:scale}) can be
generalized to deal with the critical properties of Eq.~(\ref{eq:harpnnn}).
For $t_b>t_a>t_{a\bar b}+t_{ab}$ the width for irrational $\phi$ is still
given~\cite{thouless83} by Eq.~(\ref{eq:width}), but the scaling length is
now given by
Eq.~(\ref{eq:lyap1}) so that we have
\be
 W\asymp (t_b-t_a)g\bigl( q\ln\bigl[{{1\over 2}t_b+{1\over
2}\sqrt {t_b^2-4t^2}
 \over {1\over 2}t_a+{1\over 2}\sqrt {t_a^2-4t^2}}\bigr]
\bigr) \approx
(t_b-t_a)g\bigl( q{t_b-t_a\over \sqrt{t_b^2-4t^2}} \bigr) \;.
\label{eq:scale1}
\ee
In particular, this results in the prediction that the measure of the
spectrum should scale as
\be
W\asymp {9.3299\over q} \sqrt{t_b^2-4t^2}
\label{eq:scale1c}\ee
for $t_a=t_b\ge t_{a\bar b}+t_{ab}$. A special case of this is the
triangular lattice with $t_a=t_b=t_{a\bar b}$, $t_{ab}=0$, where the
measure  of the spectrum is the same as it is for the square lattice.

For the case $t_b>2t_{a\bar b}=2t_{ab}\ge t_a$ the measure of the spectrum
for $\phi$ irrational is $4t_b-8t_{ab}$ and the scaling length is
given by Eq.~(\ref{eq:lyap2}), so the use of the same argument would lead
to a finite size scaling expression of the form
\be
W\asymp (t_b-2t_{})G\bigl(q\sqrt{t_b-2t_{ab}}{\sqrt{t_b+2t_{ab}}+
\sqrt{t_b-2t_{ab}}\over 2t_{ab}}\bigr) \;.
\label{eq:scale2} \ee
The scaling function $G(s)$ must diverge as $s^{-2}$ at the origin, so that
at the point $t_b=2t_{a\bar b}=2t_{ab}$ the measure of the spectrum is
finite, and goes to zero like $q^{-2}$ for large $q$. However, this
argument does not take account of the second length scale introduced in
Eq.\ (\ref{eq:newlength}), so we should expect the function $G$ in Eq.\
(\ref{eq:scale2}), and the coefficient of the limiting $q^{-2}$ behavior,
to depend on $t_a$ according to the form given in Eq.\
(\ref{eq:newlength}), which is periodic or nearly periodic in $q$.

The argument of Last and Wilkinson\cite{last92} which provides a lower
bound to the spectrum for the critical case can be generalized to deal with
the situations we consider in this work. The simplest case is given by
$t_b>t_a\ge t_{a\bar b}+t_{ab}$, where the result of Avron, Mouche and
Simon\cite{avron90} that the intersection spectrum (the intersection over
$k_2$ of the spectra for fixed $k_2$) has measure $4(t_b-t_a)$ remains
valid, as is shown in Appendix B. The intersection spectrum is defined, as
can be seen from Eq.~(\ref{eq:offdiag}), as the set of values of $E$ for
which $P_0(E)$ lies in the range
\be
\pm 4t^q \{ T_q({t_b\over 2t})- T_q({t_a\over 2t})\}
-(-1)^p2(t_{a\bar b}^q +t_{ab}^q) \;.
\label{eq:range1} \ee
As $t_b$ approaches $t_a$ this range approaches zero as
\be
4(t_b-t_a)t^{q-1} T_q'({t_a\over 2t})\;,
\ee
and each band centered on $E_\alpha$ has width approximately equal to this
range divided by $|P_0'(E_\alpha)|$, so that the sum rule for the derivatives
of $P$ at the points of the intersection spectrum at the critical point
$t_a=t_b\ge t_{a\bar b}+t_{ab}$ is
\be
\sum_{\alpha=1}^q {1\over |P'(E_\alpha)|} ={1\over t^{q-1} T_q'({t_a\over
2t})}\;.
\label{eq:sumrule0} \ee
For $q$ large and $t_b=t_a> t_{a\bar b}+t_{ab}$ this gives
\be
\sum_{\alpha=1}^q {1\over |P'(E_\alpha)|} \approx{\sqrt{t_a^2-4t^2}\over
2qt^q T_q({t_a\over 2t})}\approx {2^q \sqrt{t_a^2-4t^2} \over q(t_a+
\sqrt{t_a^2-4t^2})^q} \;.
\label{eq:sumrule1} \ee
For $t_b=t_a=t_{a\bar b}+t_{ab}$ it gives
\be
\sum_{\alpha=1}^q {1\over |P'(E_\alpha)|} ={t_{a\bar b}-t_{ab}\over q
(t_{a\bar b}^q-t_{ab}^q)}
\label{eq:sumrule2} \ee
for all $q$. For the bicritical case $t_b=t_a=2t_{a\bar b}=2t_{ab}$ it gives
\be
\sum_{\alpha=1}^q {1\over |P'(E_\alpha)|} ={1\over q^2 t_{ab}^{q-1}}\;.
\label{eq:sumrule3} \ee

For $t_b>t_{a\bar b}+t_{ab}>t_a$ there is no intersection spectrum, since
the ranges of the constant in Eq.~(\ref{eq:offdiag}) are nonoverlapping
for $qk_2=0$ and $qk_2=\pi$. However, we can get an exact expression for
the intersection over $k_1$ of the spectra for fixed $k_1$. This would
be the intersection spectrum for the dual problem with $t_a$, $t_b$
interchanged. This spectrum is the set of values of $E$ for which
$P_0(E)$ lies in the range
\be
\pm[4t^q T_q({t_b\over 2t})-
2t_{a\bar b}^q -2t_{ab}^q]-(-1)^p 4t^q T_q({t_a\over 2t})\;,
\label{eq:range2} \ee
and it is shown in Appendix B that the measure of this spectrum is exactly
$4(t_b-t_{a\bar b}-t_{ab})$. The same argument that led to
Eq.~(\ref{eq:sumrule0}) leads without approximation to
Eq.~(\ref{eq:sumrule2}) in the case $t_b=t_{a\bar b}+t_{ab}>t_a$, and to
Eq.~(\ref{eq:sumrule3}) in the bicritical case $t_b=2t_{a\bar
b}=2t_{ab}>t_a$.

These sum rules can be used both to generate rough estimates for the
measure of the spectrum (the union over $k_2$ of the spectrum for fixed
$k_2$) and to get rigorous lower and upper bounds, by repeating the
arguments used by Last and Wilkinson\cite{last92} and Last\cite{last93}.
The rough estimate of this sum of band widths is obtained by multiplying
the range of the constant term in the expression (\ref{eq:offdiag}) for
$P(E)$ by the appropriate expressions for $\sum 1/|P'|$ in
Eqs.~(\ref{eq:sumrule1})--(\ref{eq:sumrule3}). For  $t_b=t_a\ge t_{a\bar
b}+t_{ab}$ this gives \be
W\approx 8\sqrt{t_a^2-4t^2}/q\;.
\label{eq:approx1} \ee
For $t_b= t_{a\bar b}+t_{ab}\ge t_a$ it gives
\be
W\approx 8|t_{a\bar b}-t_{ab}|/q\;,
\label{eq:approx2} \ee
while for $t_b=2t_{a\bar b}=2t_{ab}\ge t_a$ it gives
\be
W\approx 8t_b/q^2\;.
\label{eq:approx3} \ee

The argument given by Last\cite{last93} for the upper bound needs no
modifications for the case we are considering. The result he obtained is
that if the spectrum is defined as the set of values of $E$ for which
\be
-b_1\le P_0(E)\le b_2\;,
\label{eq:spectrum} \ee
where $b_1$, $b_2$ are positive, with $P_0(E)$ a polynomial whose zeros
$E_\alpha$ are all real and distinct, and for which the zeros of the
derivative all lie outside the bands, then the sum of the widths of the
bands $W$ satisfies
\be
W<e(b_1+b_2)\sum_\alpha {1\over |P_0'(E_\alpha)|} \;.
\label{eq:ubound} \ee
The argument for the lower bound needs some modification because $b_1\ne
b_2$, but a straightforward extension of Last's argument gives
\be
W>{1\over 2}\min(b_1+\sqrt{b_1^2+4b_1b_2},b_2+\sqrt{b_2^2+4b_1b_2})
\sum_\alpha {1\over |P_0'(E_\alpha)|} \;.
\label{eq:lbound} \ee

The parameters $b_1$, $b_2$ are given by the differences between the
value of the expression in Eq.~(\ref{eq:range1}) or eq.~(\ref{eq:range2})
that defines the intersection spectrum, and the two similar
expressions  that define the band edge. For $t_b=t_a> t_{a\bar b}+t_{ab}$
we get
\be
b_{1,2}= 8t^q T_q({t_a\over 2t}) \pm 4(t_{a\bar b}^q +t_{ab}^q) \;,
\ee
and the second term in this expression becomes negligible for large
enough $q$. The bounds are therefore
\be
8e\sqrt{t_a^2-4t^2}/q >W> 2(\sqrt 5+1) \sqrt{t_a^2-4t^2}/q\;.
\label{eq:bound1} \ee

For $t_b= t_{a\bar b}+t_{ab}\ge t_a$ the parameters are
\be
b_{1,2}= 4t_{a\bar b}^q +4t_{ab}^q \pm 8t^q
T_q({t_a\over 2t})\;.
\label{eq:range3} \ee
For $t_{a\bar b}\ne t_{ab}$ only one of these terms is relevant in the
large $q$ limit, so that the bounds obtained from
Eqs.~(\ref{eq:sumrule2}), (\ref{eq:ubound}), (\ref{eq:lbound}) and
(\ref{eq:range3}) are
\be
8e|t_{a\bar b}-t_{ab}|/q >W>  2(\sqrt 5+1)|t_{a\bar b}-t_{ab}|/q\;.
\label{eq:bound2} \ee

For $t_b=2t_{a\bar b}=2t_{ab}\ge t_a$ the $t_a$ dependent term in
Eq.~(\ref{eq:range3}) is also relevant, and
Eqs.~(\ref{eq:sumrule3}, (\ref{eq:ubound}), (\ref{eq:lbound}) and
(\ref{eq:range3}) give
\be
{8et_b\over q^2}> W> {2t_b\over q^2}\sqrt{1-\gamma}[\sqrt{5+3\gamma}
+\sqrt{1-\gamma}] \;,
\label{eq:bound3} \ee
where
\be
\gamma= |T_q({t_a\over 2t_{ab}})|
\ee
This is an example of the importance of the second relevant length scale
mentioned in connection with Eq.~(\ref{eq:newlength}).

For the case $t_a=t_b=2t_{a\bar b}=2t_{ab}$ this line of argument gives us
no useful lower bound, since all our lower bounds reduce to zero. For
$t_{a\bar b}+t_{ab}>t_b\ge t_a$ we have not succeeded in finding an exact
expression for the intersection spectrum or some equivalent spectrum.

\section{Multifractal analysis }

In this Section, we perform  numerical scaling analyses for the energy
spectrum when $\phi=p/q$ approaches the quadratic irrational number $1/\tau
\equiv (\sqrt{5}-1)/2$: the inverse of the golden mean.  We take $p/q$ to be
$F_{n-1}/F_n$, where $F_n$ is the $n$th Fibonacci number defined recursively
by $F_n = F_{n-1}+F_{n-2}$ and $F_0=F_1=1$.  Note that $q=F_n \sim \tau^n$
for a large $n$. In order to obtain a spectrum for  a rational approximant,
the Bloch theorem is applied first to Eq.~(\ref{eq:nnn}), which is
periodic with period $F_n$. Then the system becomes effectively finite
and the spectrum  is obtained by a numerical diagonalization.

To discuss localization of the eigenstates of Eq.~(\ref{eq:nnn}) we examine
its spectrum for fixed $k_2$. When $\phi = p/q$, the spectrum consists of
$q$ bands whose widths are denoted by $\Delta_i$ ($i=1, \ldots , q$). Since
each band has the same number of states, we assign a probability measure 1/q
to each band. With increasing $q$, each band splits into many sub-bands. In
order to understand the scaling of the spectrum, we introduce a scaling
index $\alpha$ by \be
 1/q \sim \Delta_i^{\alpha_i}\; .
\label{eq:defeps}\ee
If states are localized
in the limit of $q\rightarrow\infty$, the band widths $\Delta$ decrease
exponentially with $q$, so that $\alpha$ goes to zero. This corresponds to a
point spectrum. If states are extended, on the other hand, $\Delta$ scales as
$1/q$, so $\alpha$ is $1$. This corresponds to an absolutely continuous
spectrum. In the critical case, $\alpha$ is expected to take  values between
0 and 1. The values of $\alpha$ have a distribution on the whole spectrum.
This situation corresponds to a singular continuous spectrum.

It is clear from the analysis of the difference between the union
spectrum and the intersection spectrum given in Sec.\ III\cite{last92}
that the value of $k_2$ only plays a crucial role for the discrete
spectrum, when the eigenstates are localized. For the absolutely
continuous spectrum there is only an exponentially small dependence on
the value of $k_2$. For the critical case, since the intersection
spectrum of zero measure divides the union spectrum into two equal
halves, it is clear that to take fixed $k_2$ gives a spectrum whose
measure is roughly half that of the union spectrum. In the subsequent
discussion we will avoid the region of localized states, and consider
the spectrum found by taking the union over all values of $k_2$.

For a systematic analysis of  systems with such complex scaling behavior, it
is convenient to use the multifractal technique developed by Halsey {\it et
al.}\cite{halsey86} They have introduced the spectrum of singularity
$f(\alpha )$ defined by
\be
 \Omega (\alpha ) \sim \langle\Delta\rangle^{-f(\alpha )}\; ,
\ee
where $\Omega (\alpha )d\alpha$ is a number of bands whose scaling index
$\alpha$ lies between $\alpha$ and $\alpha +d\alpha$,
and $\langle\Delta\rangle$ is a representative value of $\Delta$ which was
not specified clearly in Ref. 23. We first explain the multifractal
technique as reformulated  by Kohmoto.\cite{kohmoto88}

\subsection{Formulation}

First introduce
a scaling index $\epsilon$  by
\be
 \epsilon_i = -\frac{1}{n} \ln \Delta_i\; .
\ee
It is related to $\alpha$ by
\be
 \alpha\epsilon = \ln \tau.
\label{eq:relalpeps}\ee
We also define an ``entropy function'' $S(\epsilon )$ by
\be
 S(\epsilon ) = \frac{1}{n} \ln \Omega '(\epsilon ) ,
\label{eq:defent}\ee
where $\Omega '(\epsilon )d\epsilon$ is the number of bands
whose scaling index lies between $\epsilon$ and $\epsilon +d\epsilon$,
namely $\Omega '(\epsilon ) = \Omega (\alpha )|d\alpha /d\epsilon |$.
Here it is important to notice that $\Delta_i$
and $\Omega '(\epsilon )$ depend exponentially on $n$.
A band at the $n$th level splits into many
bands at a higher level and may thus yield a number of different values
of the scaling indices $\epsilon$.  However, we expect that the entropy
function which represents the distribution of $\epsilon$ will converge to
a smooth limiting form as $n$ tends to infinity, and give complete
information about the scaling behavior. As in the formalism of statistical
mechanics, it is convenient to introduce a ``partition function'' and a
``free energy'' which are defined by
\be
 Z_n(\beta) = \sum_{i=1}^{q} \Delta_i^{\beta}
\label{eq:defpart}\ee
and
\be
 F(\beta ) = \lim_{n\rightarrow\infty } \frac{1}{n} \ln Z_n(\beta ).
\label{eq:deffree}\ee
Once the free energy is calculated, the entropy function is
obtained by a Legendre transformation,
\be
 S(\epsilon ) = F(\beta ) + \beta\epsilon ,
\label{eq:lgtr}\ee
\be
 \epsilon = - \frac{dF(\beta )}{d\beta } .
\ee
Thus by changing ``temperature'' $\beta$ one can pick a value of $\epsilon$
and then the corresponding $S(\epsilon )$ is calculated.
On the other hand, $\beta$ can be written in terms of $\epsilon$ as
\be
\beta = \frac{dS(\epsilon )}{d\epsilon } .
\label{eq:beta}\ee
Usually $S(\epsilon )$ is defined on an interval
$[\epsilon_{\mbox{\small min}}, \epsilon_{\mbox{\small max}}]$
and there is no scaling behavior corresponding to $\epsilon$ which is
outside the interval and $S(\epsilon ) = 0$.  However, $F(\beta )$
is still defined there and from (\ref{eq:lgtr}) it is given by
$F(\beta ) = -\epsilon_{\mbox{\small max}}\beta$
for $\beta > \beta_{\mbox{\small min}}$ and
$F(\beta ) = -\epsilon_{\mbox{\small min}}\beta$
for $\beta < \beta_{\mbox{\small max}}$.  Thus useful
information is only contained in $F(\beta )$ for the region between
$\beta_{\mbox{\small min}}$ and $\beta_{\mbox{\small max}}$
where it is not linear.

Using Eq.(\ref{eq:relalpeps}) and identifying
$\langle\Delta\rangle = \exp (-n\epsilon )$
(see (\ref{eq:defeps})), $f(\alpha )$ is related to
the entropy function by
\be
 f(\alpha ) = \frac{S(\epsilon )}{\epsilon } .
\label{eq:fa}\ee
$f(\alpha )$ is defined on an interval $[\alpha_{\mbox{\small min}},
\alpha_{\mbox{\small max}}]$ where
$\alpha_{\mbox{\small min}}=\ln\tau/\epsilon_{\mbox{\small max}}$
and $\alpha_{\mbox{\small max}}=\ln\tau/\epsilon_{\mbox{\small min}}$.
 The maximum value of $f(\alpha )$ gives the Hausdorff dimension of the
spectrum.

\subsection{Numerical results for $f(\alpha )$ }

In all our numerical work we have taken $t_{a\bar b}=t_{ab}$, so we
refer just to the value of $t_{ab}$ in the rest of this Section.
Since we are studying the union of the spectrum over all values of $k_2$, the
system is symmetric with respect to  an interchange of $t_a$ and $t_b$.
Thus we carry out numerical calculations only for
$t_b \le t_a$.  The system is characterized
by two  parameters $t_b/t_a$ and $t_{ab}/t_a$.

Before going to the numerical results we notice that the results
of Sec.\ III for the Lyapunov exponents
(in particular Eq.~(\ref{eq:lyap1}) and
the discussions below it) determine types of spectra
in some parts of the parameter space. In region I in Fig.~\ref{fig:hiramoto1}
($t_{ab}/t_a < 1/2$), the eigenstates are extended in the $x$ direction
for irrational $\phi$. Thus the spectrum is
absolutely continuous and the minimum of $\alpha$ is
$\alpha_{\mbox{\small min}} =1$
and $f(\alpha_{\mbox{\small min}})=1$. On line  $BC$ ($t_a = t_b$),
Eq.~(\ref{eq:nnn}) is self-dual and the Lyapunov exponents
for both $x$ and $p_x$ directions are zero. Then
the spectrum is expected to be singular continuous.The properties of region
II have not been determined by the analysis of the Lyapunov exponents in
Sec.\ III, but it will  be shown numerically that the spectrum is singular
continuous in the whole region II.

The energy spectra for $t_a = t_b$ (self-dual line $BD$)
and $n=10$ are shown in Fig.~\ref{fig:hiramoto2}(a).
As is well known, the spectrum for $t_{ab}=0$ (critical point of
pure Harper's equation) has a self-similar structure.
The whole spectrum has three main bands, each main band has three subbands,
and so on. This structure remains unchanged
 for $0<2t_{ab} < t_a$, but at $2t_{ab}=t_a$ it changes.  For example,
band edges $a$ and $b$ in pure Harper's equation in Fig.~\ref{fig:hiramoto2}(a)
continuously change to $a'$ and $b'$ as $t_{ab}$ is increased. They are
still band edges for $2t_{ab}<t_a$. When $2t_{ab}\rightarrow t_a$, however,
the two points
tend to the same point $c$ and are no longer band edges.  See
Fig.~\ref{fig:hiramoto2}(b). For $2t_{ab} > t_a$, the topological structure
changes further.

In order to see the global scaling behavior, we have performed numerical
 calculations of $f(\alpha )$, which is defined in the limit of
$n\rightarrow\infty$, by extrapolating the numerical
data up to $n=19$ ($q=F_{19}=6765$).

Figure \ref{fig:hiramoto2} is a plot of $f(\alpha)$ for $t_{ab}=.4$. This plot
is identical, within the precision of the plot, to the plot obtained for
the case $t_{ab}=0$ (pure Harper's equation). As was pointed out by
Tang and Kohmoto\cite{tang86}, $f(\alpha )$ is defined on the interval
between $\alpha_{\mbox{\small min}}\simeq 0.421$
and $\alpha_{\mbox{\small max}}\simeq 0.547$, and takes the maximum value
$f=0.5$ at $\alpha =0.5$. Calculations for a variety of other values of
$t_{ab}/t_a$ in the range $0<t_{ab}/t_a < 0.5$ give results that look
identical to Fig.~\ref{fig:hiramoto3}, and we conclude that $f(\alpha )$ is
universal in this range. At the critical point of pure Harper's
equation, $\alpha_{\mbox{\small min}}$ is the scaling index of the edges
of the spectrum (and the edges of each subband), whereas
$\alpha_{\mbox{\small max}}$ is the scaling index of the center of the
spectrum (and the center of each subband). Even if $t_{ab}$ is non-zero,
the situation remains the same as long as $2t_{ab}<t_a$. More
specifically, the scaling index of each band is identical to that of the
topologically corresponding band of pure Harper's equation.

At the bicritical point $C$ ($2t_{ab}=t_a=t_b$) the shape of $f(\alpha )$
suddenly changes. The nonzero values of $f$ are found in the range
between $\alpha_{\mbox{\small min}}\simeq 0.272$ , $\alpha_{\mbox{\small
max}}\simeq 0.421$, and the maximum value of $f$ is about 0.33 at
$\alpha\simeq 0.33$. Note that $\alpha_{\mbox{\small max}}$ at this point
is identical to $\alpha_{\mbox{\small min}}$ for $2t_{ab} < t_a$.
At $2t_{ab}=t_a$, however, the topological structure of the spectrum
and the scaling indices are different.
For example, the scaling index of the bands coming from
the centers of the subbands of pure Harper's equation is 0.303.
It is 0.289 at the edges of bands ({\it e.g.}, $d$, $e$
and $f$ in Fig.~\ref{fig:hiramoto2}(b)).  On the other hand, the index remains
0.421 at $c$ in Fig.~\ref{fig:hiramoto2} and becomes $\alpha_{\mbox{\small
max}}$.

Plots of $f(\alpha )$ on line $CD$ (  $2t_{ab} > t_a=t_b$) were obtained
for some higher values of $t_{ab}/t_a$. Although $f(\alpha )$ is not
universal on line $CD$, there is some similarity between the curves we
found, in that the maximum points are nearly at $\alpha= 0.4$
and  the maximum values are also about 0.4. The values of $\alpha_{\mbox{\small
min}}$, $\alpha_{\mbox{\small max}}$, $\alpha_0$, the position of the
maximum of $f$, and $f(\alpha_0)$, the maximum value, are shown in Table
I.

We have also calculated the form of $f$ on the line $AC$
($2t_{ab}/t_a=1, t_a > t_b$). As in the previous case on line $CD$, the
curves are somewhat similar to one another in the sense that the maximum
points are at $0.3 < \alpha < 0.35$ and the maximum values are also
between 0.3 and 0.35. The main features of these results are also given
in Table I.

Finally we have calculated $f(\alpha )$ in region II, and one example
is given in Table I. In this region, the convergence of the
extrapolation  $n\rightarrow \infty$ from the numerical data for finite
$n$'s is not so good as the previous cases. Thus
we cannot obtain reliable results of $f(\alpha )$
to claim some universal features. Even though the errors are rather large,
however,  it appears that $f(\alpha )$ is a continuous function defined on
a finite range of $\alpha$. Thus we conclude that  the spectrum
is multifractal and singular continuous in region II.

\section{Numerical results for band widths}

\subsection{Total band width for Fibonacci sequence}

In the previous Section, we have obtained the following:
(1) for $t_a = t_b$ and $2t_{ab} < t_a$ (line $BC$ in
Fig.~\ref{fig:hiramoto1}),
the scaling behavior of
the spectrum is universal, that is,  $f(\alpha )$ is completely identical to
that of the $t_{ab}=0$ case ( pure Harper's equation);
(2) at the bicritical point of $t_a=t_b$ and $2t_{ab}=t_a$
(point $C$ in Fig.~\ref{fig:hiramoto1}),
the scaling behavior suddenly changes; (3) when $t_a = t_b$ and
$2t_{ab} > t_a$ (line $CD$ in Fig.~\ref{fig:hiramoto1}), the scaling behavior
is clearly different from
those of (1) and (2). Although it is not completely universal,
the changes are small for increasing $t_{ab}/t_a$;
(4) when $t_a > t_b$ and
$2t_{ab} = t_a$ (line $AC$ in Fig.~\ref{fig:hiramoto1}), the
scaling behavior is similar but not identical to that of point
$C$ ($t_a = t_b$ and $2t_{ab}=t_a$); (5) when $2t_{ab} > t_a$ and
$t_a > t_b$ (region II in Fig.~\ref{fig:hiramoto1}), it is difficult to
estimate $f(\alpha )$
but it is certain that the spectrum is also multifractal
and singular continuous.

To make the above statements more concrete, we investigate scaling of
the total band widths. Recall that in  pure Harper's equation in
the critical case ($t_a = t_b, t_{ab}=0$), the total band width $W$
scales as $W \sim 1/q$ for large $q$ (see Eq.\ (\ref{eq:scale})). We
know from the analytic results of Sec.\ III that this scaling also
holds all along the line $BC$ ($t_a=t_b>2t_{ab}$) except at $C$. The
numerical results show that for the Fibonacci sequence the result
\be
qW\asymp 9.3299 \sqrt{t_a^2-4t_{ab}^2}
\ee
holds accurately, in agreement with Eq.\ (\ref{eq:scale1c}).

On the bicritical line $AC$ ($2t_{ab}=t_a\ge t_b$), which separates the
region of $t_a$ dominant from the region of $t_{ab}$ dominant, we know
from the results of Sec.\ III that the sum of the band widths should
scale as $q^{-2}$. Figure ~\ref{fig:hiramoto4} shows the results for $q^2W$
plotted against $n$ ($\sim \ln{q}/\ln\tau$) for different values of $t_b/t_a$
on this line. In this subsection all the plots of $W$ are in the units such
that  $t_a = 1$.  It is seen that the scaling index $\delta$ for the
global behavior of $W$ which is defined by
\be
 W \sim (1/q)^{\delta}\; ,
\ee
is 2. For $t_a=t_b=2t_{ab}$, the quantity $WF_n^2/t_a$ tends rapidly to
6.4911.   In addition to this power-law decrease, an oscillatory
behavior is observed for $t_b\ne t_a$. The oscillation has a
period 4 against $n$ in the case $2t_b=t_a$, which is probably related
to the period 8 of the Fibonacci sequence modulo 3.

In the region $2t_{ab}>t_a\ge t_b$ we find two different types of
behavior, but in all cases the analysis is complicated by oscillatory
terms superposed on a general power law dependence on $q$. Plots of
$\ln W$ against $\ln F_n$ (which is proportional to $n$) appear to lie
close to a line of slope $-1.25$ for the case $2t_{ab}>t_a= t_b$ (the
line $CD$ in Fig.~\ref{fig:hiramoto1}), whereas they cluster around a line of
slope $-1.56$ for the case $2t_{ab}>t_a> t_b$ (the interior of the region II
in Fig.~\ref{fig:hiramoto1}). Figure~\ref{fig:hiramoto5} shows plots of
$\ln(F_n^{1.25}W)$ against $n$ for various examples of $2t_{ab}>t_a=t_b$.
There is a period 4 oscillation in the case
$t_{ab}=t_a=t_b$, similar to the oscillation of period 4 that shows up in
Fig.~\ref{fig:hiramoto4} for the case $t_{ab}=t_a/2=t_b$. There is also an
oscillation of period 6 for the case $t_{ab}=t_a/\sqrt 2$. In other cases the
oscillation around the general horizontal trend is irregular, and show no
signs of diminishing as $n$ increases. An accurate estimate of the exponent
$\delta$ can be made when the period is short, but we can only make a rough
estimate when there is no period within the range of $n$ we can use. There
appears to be a slight difference, of order .01, between the values we get
for $\delta$ with $t_{ab}=t_a=t_b$ and with $t_{ab}=t_a/\sqrt 2=t_b/\sqrt 2$.

Figure ~\ref{fig:hiramoto6} shows plots of $\ln(F_n^{1.56}W)$ against $n$ for
various examples of $2t_{ab}>t_a> t_b$. Again, there are large oscillations
about the general horizontal trend of the plots, and a simple period,
6 in this case, can be seen clearly for $t_{ab}=t_a/\sqrt 2$, $t_b=0$.
For reasons which are discussed later in this paper, we think that
this period is related to the period 12 of the Fibonacci sequence
modulo 4. The value of $\delta=1.56$ is not so easy to estimate for
these examples, but it is clearly intermediate between the value we
found on the line $CD$ of Fig.~\ref{fig:hiramoto1}, and the value of 2 which
is known to be correct for the bicritical line $AC$.

The results in this subsection are summarized as follows.
The scaling index $\delta$ for the total band width is 1 on
the critical line $BC$ in Fig.~\ref{fig:hiramoto1}. Not only at point $C$ but
also on line $AC$, $\delta$ is 2, {\sl i.e.}, the bicritical
behavior is found.  On line $CD$, we cannot find a significant
dependence of $\delta$ on $t_{ab}/t_a$, and it is
about 1.25.  Also in region II, the situation is similar and
the value of $\delta$ is about 1.56.

In the limit of $t_{ab}/t_a \rightarrow\infty$, the problem reduces to
the case only with nearest neighbor couplings. Thus the $\delta$ must be 1 in
this limit.  Although we investigated $\delta$ for quite large $t_{ab}/t_a$
(up to $t_{ab}/t_a = 10$) in region II and on the line $CD$, we could not
find a tendency that $\delta$ decreases and approaches 1.

\subsection{Scaling for $\phi=1/q$}

For rational approximants to the golden mean the total band width is
spread over a large number of bands in different energy ranges. For large
denominator rational approximants to a small denominator rational, say
$p_0/q_0$, the total band width is concentrated in narrow ranges about
the $q_0$ values of the energy where there is a logarithmic singularity
in the density of states for the case $\phi=p_0/q_0$.
\cite{last92,thouless91b}
For sequences such as $\phi=1/q$ or $2/q$ the limit of the sequence gives
$p_0=0$, $q_0=1$, and all the width comes from one singular energy at or
near the center of the band. To a considerable extent the results for
this case can be understood in terms of the WKB analysis presented in
Sec.\ VI, and many, but not all, of the results appear to generalize to
more complicated sequences of fractional values of $\phi$.

We have made some numerical checks of the scaling relations
(\ref{eq:scale}) and (\ref{eq:scale1}) for the case
$t_b>t_a>2t_{ab}$, and of the scaling relation (\ref{eq:scale2}) for
$t_b>2t_{ab}>t_a$, but these have not been extensive, and we have not
found particularly interesting features. We have concentrated on
three critical or bicritical cases. For $t_b=t_a>2t_{ab}$ the bounds
(\ref{eq:bound1}) show that the width must scale as $1/q$, and
finite size scaling theory suggests that it should have the limiting
form given by Eq.~(\ref{eq:scale1c}). For the bicritical case
$t_b=2t_{ab}\ge t_a$ the bounds Eq.~(\ref{eq:bound3}) show that the width
must scale as $1/q^2$ (except possibly in the special case $t_a=t_b$),
but Eq.~(\ref{eq:scale2}) does not tell us much more, since we expect
the additional length given by Eq.~(\ref{eq:newlength}) to be involved.
Finally, there is the critical case $2t_{a\bar b}=2t_{ab}>t_b\ge t_a$,
which one might expect to be analogous to the other critical case, since
the dominant terms are essentially the same, but rotated through an angle
$\pi/4$ in the $p_x,x$ plane, with half the size of unit cell. However,
we know no rigorous bounds in this case, have derived no useful finite
size scaling relations, and expect the influence of the additional
lengths given by Eq.~(\ref{eq:newlength}) to be important.

The case $t_a=t_b>2t_{ab}$ is critical and the measure $W$ scales like
$\sqrt{t_a^2-4t_{ab}^2}/q$ as predicted by the finite size scaling theory
of Eq.~(\ref{eq:scale1c}). We have done a numerical check for numerator 2
and $q$ odd,  where we have evaluated $qW/ \sqrt{t_a^2-4t_{ab}^2}$ for
$t_{ab}$=0, 0.2, 0.4 and found that they very rapidly converge to a
common value. The difference between values of
$qW/\sqrt{t_a^2-4t_{ab}^2}$ for different values of  $t_{ab}$ is less than
one part in $10^5$ for $q$ greater than 41.  This implies that the energy
scale is reduced by a factor  $\sqrt{1-4t_{ab}^2/t_a^2}$ and that there
are no logarithmic corrections in the case of $p$=2 as we increase
$t_{ab}$.

For numerator equal to unity, the scaling limit remains the same, but
there are large corrections to
scaling, which are shown in Fig.~\ref{fig:han1}. These show an interesting
cusp-like oscillation of $qW/\sqrt{1-4t_{ab}^2/t_a^2}$ for non-zero
$t_{ab}$. The periodicity of this oscillation can be related to the
integral of the classical momentum over the length of the system, as we
discuss in Sec.\ VI, while the magnitude of the oscillation is bounded by
two  curves which are followed by odd and even values of $q$ with
$t_{ab}$=0.\cite{thouless91a}

The $2t_{ab}>\max(t_a,t_b)$ region is analogous to the $t_a=t_b$ dominant
region just considered in that, without $t_a,t_b$, the problem reduces to
that  of the Harper's equation with twice as much flux per unit cell, and
that  $t_a,t_b$ may be regarded as perturbations to the critical Harper
problem. It is probably not important that the perturbations generally
break the square symmetry of the  system, since, as we discussed in Sec.\
II, the symmetry under $p_x\to -p_x$ remains, as well as rotation of
phase space by $\pi$. In the cases we have examined with $\phi=1/q$ the
measure scales like $1/q$ for $\max(t_a,t_b)<2t_{ab}$. As was mentioned in
Sec.\ III,  $qW$ does not seem to tend to a limit, but oscillates in the
$t_{ab}$  dominant regime.

To study the periodic nature of the measure, it is convenient to choose
the  set of parameters $t_a/2t_{ab}=\cos (\pi p_1/q_1)$, $t_b/2t_{ab}=\cos
(\pi p_2/q_2)$, and characterize the system with a set of fractions
$(p_1/q_1,p_2/q_2)$. Several graphs of $qW$ as a function of $q$ are shown
in Fig.\ \ref{fig:han2}.  For $p_1=p_2=1$, we have found strong peaks at
multiples of $q_1\times q_2$ (primary peaks) and much less strong peaks at
multiples of $q_1$ (secondary peaks)  if $q_1$ is a small integer such as 2
or 3. For $q_1$ and $q_2$ both  large,\,close,\, and relatively prime, for
example, 5 and 6 or 5 and 7,  the secondary peaks do not seem to occur at
definite multiples of either number. The primary peaks always occur at
multiples of $q_1\times q_2$.  If $q_1=q_2$ or they share a common factor,
then the secondary peaks occur at integer multiples of their lowest
common multiple, and the values of $qW$ at these points nearly
match those of the primary peaks.

In the bicritical case with $2t_{ab}=t_b>t_a$ we know from the
inequalities (\ref{eq:bound2}) that the measure of the spectrum must
be proportional to $q^{-2}$, and there is one length scale remaining
which is given by Eq.\ (\ref{eq:newlength}), so we might expect a scaling
form
\be
q^2 W=g_{bc}(q\arccos(t_a/2t_{ab}))\;.
\label{eq:scale3}\ee
Figure \ref{fig:han3} shows $g_{bc}(x)$, with $x=q\arccos(t_a/2t_{ab})$,
for $t_a/2t_{ab}$=1/3 and 0.99. In the limit $t_a=2t_{ab}$, $g_{bc}(x)$
becomes a slowly increasing function of $x$, and we have not determined
its limiting value; this slow convergence may be a special feature of
particular fractional forms of $\phi$, as we found no sign of it for
the Fibonacci sequence. Figure \ref{fig:han3}(b) transparently displays a
cusp-like form of the scaling function, whose variation lies well within
the bounds given by Eq.~(\ref{eq:bound3}).  Earlier in this subsection,
such cusps were reported when we considered the $t_a=t_b$ dominant region. It
turns out that a function of the form  $g_{bc}(x)=
A-B\ln(1+|\sin(\pi(x-\delta)|)$, where $A,B$ and $\delta$ are constants
chosen to fit, describes the actual scaling function rather well. The
motivation for this form of $g_{bc}(x)$ came from the idea that $x$ must
be a dimensionless quantity and that such a quantity can be obtained by
multiplying $q$ on both sides of Eq.\ (\ref{eq:lyap}) {\it before} one
takes the limit $q\to\infty$. Since $t_{a\bar b}=t_{ab}$, this gives us
$x=q\lambda_x=\ln2-\ln(1+|T_{q}(t_a/2t_{ab})|)$ which seems to contain
essential features of $g_{bc}(x)$. Based on this hypothesis, the cusp-like
behavior can be understood as the reflection of the importance of the
absolute value of the Chebyshev polynomial.

\subsection{Scaling for $\phi=p/(p^2\pm 1)$}\label{bw3}

The sequences $\phi=p/(p^2\pm 1)$ are intermediate between
$\phi=1/q$ and $\phi=F_n/F_{n-1}$. The continued fraction expansion has
only two terms in it, rather than the $n/2$ terms for the Fibonacci
sequence, and it represents a slow modulation of the diagonal term, and so
could be treated by WKB methods (although we have not succeeded in
carrying out such an analysis in this case). It is, however, more
complicated than the $\phi=1/q$ case in that the significant
contributions to the band width come not only from bands in the
immediate vicinity of the singular energy $-t_at_b/t_{ab}$, but also
from the centers of neighboring clusters of $p$ subbands.

We have studied the bicritical case $t_b=2t_{ab}$=1 and $t_a=
\cos(p_1\pi/q_1)$ where  $q_1$ was kept small ($\leq$ 7). With this
choice of $t_a$ we expect behavior periodic in $q$ in the limiting values
of $q^2W$ coming from the constant term Eq.\ (\ref{eq:offdiag}), but
there may also be some dependence on the numerator $p$ coming from other
terms in the characteristic equation. We have confined ourselves to
numerators not  exceeding 20,\, and denominators up to 401. As in the
previous case of a simple fraction of $\phi$, the scaling function  showed
periodicity (up to corrections to scaling) with periods equal to $q_1$ for
both $p^2\pm 1$. The pattern of graphs are quite different in two cases and
values of maxima and minima as well as where the maxima and minima occur
do not agree in general. The convergence to the limit was much slower here
than it was with  $\phi=1/q$. This might have to do with the fact
that, for the
fraction  $\phi=p/(p^2\pm 1)\sim 1/p$ with $p\leq$ 20, the
corrections are not yet completely negligible.  For $p_1/q_1=1/4$, we
should expect a period of 2 because $p^2\pm 1\pmod 4$ alternates between
2(0) and 1(3) for odd and even $p$ but instead we observe  a period of 4.
Apparently the scaling function here is more complicated than  what was
the case if the relevant variable was simply the ratio of two length
scales of the system. This shows conclusively that there is periodic
dependence on the value of the numerator as well as on the value of the
denominator.

In the $t_{ab}$ dominant critical region we have found instances of an
anomalous (non-integer) scaling exponent. Figure \ref{fig:han4} shows a
plot of $\log (qW)$ against $\log q$ for the case $2t_{ab}=1, t_a=0$,
$t_b=\cos (\pi/4)$, $\phi=p/(p^2+1)$. For numerators equal to $2\bmod
4$ we
see the points lying on a line with a negative slope, whereas  linear
scaling should give no slope at all, as we see for the other values of
$p$. The slope both of this plot, and of the very similar plot for
$\phi=p/(p^2-1)$, is found to be within 0.3\% of -1/2, which implies
that, for the sequence $p/q=2(2n+1)/[4(2n+1)^2\pm 1]$,  the scaling is
like $1/q^{3/2}$ rather than $1/q$. For $t_a=0,t_b= \cos(\pi/3)$, the plot
alternates between points with exponent  close to 1, if $p=0,\pm 1
\pmod 6$, and those with exponent close to  $1.5$ if $p=2,3,4\pmod 6$
for both types of denominator. It is much  harder to extract the
critical exponents here because we have to  increase $p$ by 6
instead of 4 to arrive at points lying on the same line. For
$0<t_a=t_b<2t_{ab}$,  the evidence for non-integer exponent is far less
obvious and we are not sure  if the exponent is significantly different
from 1.

The noninteger exponents found in our studies of the Fibonacci sequence
are likely to be related to these results.

\section{WKB theory}

The saddle point value of the energy contour $E_s$ corresponds to quantum
mechanical states that can thread the system without attenuation and there
exists an interesting
relation between the integral of the classical momentum $p_x$ given by Eqs.
(\ref{eq:cospx})-(\ref{eq:contour2}) between turning points with
periodicities in the scaling functions.

In Ref.\ \onlinecite{thouless90} it was shown that for $\phi=1/q$ the sum
of the band widths could be related to the Green function at the turning
points. For $t_a=t_b>t_{a\bar b}+t_{ab}$, energy close to
$-2(t_{a\bar b}+t_{ab})$ and $x=\phi/2+\xi$, where $\xi$ is small, the
continuum approximation for Eqs.\ (\ref{eq:harpnnn}) and (\ref{eq:nnn})
takes the form
\[
H +2(t_{a\bar b}+t_{ab}) \approx
\]
\be
4\pi^2\phi^2(t_a+t_{a\bar b}+t_{ab}){
d^2\over d\xi^2} +(t_a-t_{a\bar b}- t_{ab})\xi^2 +2\pi i\phi (t_{a\bar
b}-t_{ab})(\xi{d\over d\xi}+ {d\over d\xi}\xi)\;.
\label{eq:continuum1} \ee
A very similar expression is obtained near the other turning point
with $x$ close to zero and $p_x$ close to $\pi$. This can be diagonalized
by a canonical transformation, and the energy scale it yields is
\be
4\pi\phi\sqrt{t_a^2-4t_{a\bar b}t_{ab}}
\ee
in both cases. The earlier
arguments\cite{thouless90,thouless91a,thouless91b,last92} applied to this
case give the scaling result quoted in Eq.\ (\ref{eq:scale1c}).

The corrections to this scaling form can be calculated by using the WKB
approximation to get a more precise approximation to the band
widths\cite{thouless91a}, which involves the connection between the
turning points as well as the behavior at the turning points. We have not
worked this case out in detail, but we know that for $2t_{a\bar
b}=2t_{ab}$ the phase change around an orbit close to the critical orbit
is
\be
{1\over 2\pi\phi}\oint p_x dx\approx {2\over \pi\phi} \int_{0}^{\pi}dx
\arccos\left(-{2t_{ab}+t_a\cos x \over t_a+2t_{ab}\cos x}\right)\;.
\label{eq:action}\ee
Differentiation of the right side of this equation with
respect to $t_{ab}$ gives an integral that can be evaluated
explicitly, and reintegration of this result gives
\be {q\over 2\pi}\oint p_x dx \approx q\pi-{4q\over \pi}
\int_{0}^{\tanh^{-1}(2t_{ab}/t_a)} {\theta\over\sinh\theta}d\theta\;.
\ee
The first term in this expression gives rise to the correction to scaling
that alternates with the parity of $q$ even for $t_{ab}=0$, while the
second term gives a period in $q$ that goes like $\pi^2t_a/4t_{ab}$ for
small $t_{ab}$. For $t_{ab}/t_a=0.1$ it gives 24.5 as the period, which
is in good agreement with the numerical results shown in Fig.\
\ref{fig:han1}. In the limit $2t_{ab}\to t_a$ the second term is $\pi q$
and cancels the periodicity due to the first term.

For $2t_{a\bar b}=2t_{ab}$ dominant the problem is somewhat different. The
classical contours through the saddle points are given by
\be
{1\over t_{ab}}(t_b+2t_{ab}\cos p_x)(t_a+2t_{ab}\cos x) =0\;.
\label{eq:contour3} \ee
The quadratic approximation to the Hamiltonian near one of the saddle
points is given by
\be
H+{t_at_b\over t_{ab}}\approx \pm {2t_{ab}\over 2\pi\phi}i
\sqrt{(1-{t_a^2\over 4t_{ab}^2}) (1-{t_b^2\over 4t_{ab}^2})}\; ( \xi{d\over
d\xi}+{d\over d\xi}\xi) \;,
\ee
and diagonalization of this by a canonical transformation gives an energy
scale
\be
4\pi\phi t_{ab} \sqrt{(1-{t_a^2\over 4t_{ab}^2}) (1-{t_b^2\over
4t_{ab}^2})} \;.
\label{eq:saddlepoint2} \ee
This quantity gives a good account of the relative sizes of the energy
scales of the band widths shown in Fig.\ \ref{fig:han2}. However, as we
showed in the study of the characteristic equation in Sec.\ III, the
lengths given in Eq. (\ref{eq:newlength}) are certainly relevant, and
there must be terms periodic or nearly periodic in $q$. These should come
from the rectangular contours given in Eq.\ (\ref{eq:contour3}), whose
areas are
\be
4\arccos(\pm{t_a\over 2t_{ab}})\arccos(\pm{t_b\over 2t_{ab}})\;,
\label{eq:areas} \ee
and it is the ratio of these areas to the quantum of action $4\pi^2\phi$
given by Eq.\ (\ref{eq:pxdef}) that determines this periodicity.

Qualitatively this accounts for the periods in $q$ of 9 and 21 which
show up in Figs.\ \ref{fig:han2}(a) and (b), since the smallest
areas given by Eq.\ (\ref{eq:areas}) are $4\pi^2/9$ and $4\pi^2/21$
in the two cases, and the larger rectangles are multiples
of these. To understand these results in more detail we need to make a
more careful study of the way the phases affect the dynamics.

The bicritical case $t_b=2t_{ab}$ is simpler, since one of the arccosines in
Eq.\ (\ref{eq:areas}) is equal to $\pi$. For $t_a=t_{ab}$ the two
rectangles have areas $4\pi^2/3$ and $8\pi^2/3$, so the period 3 in $q$
which can be seen in Fig.\ \ref{fig:han3}(a) should be expected, while
for $t_a=.495t_{ab}$, $\pi/\arccos(t_a/2t_{ab})$ is equal to 22.2, which agrees
well with the period shown in Fig.\ \ref{fig:han3}(b).

For a large denominator fraction $p/q$ approximating a small denominator
rational $p_0/q_0$, the commutator $[x,p_x]$ remains finite and the WKB
approach does not directly apply. Wilkinson\cite{wilkinson87} has shown that
at $t_a=t_b, t_{ab}=t_{a\bar b}=0$, each one of $q_0$ clusters of bands are
described by an effective Hamiltonian obtained from quantizing the
inverse of the characteristic polynomial\ (\ref{eq:offdiag}) for the $n$th
band,
\be
E_n(k_1,k_2)=P^{-1}_{0,n}(2t_a^{q_0}\cos(q_0k_1)+2t_b^{q_0}\cos(q_0k_2))\;.
\ee
The new flux for each cluster depends on the Chern number for that
particular band, but is in general of the order of the difference
$p/q-p_0/q_0$, therefore small.
A more direct way\cite{thouless91b} is to regard the problem
as having $\phi=p_0/q_0$ with the wavevector $k_2$ slowly modulated with a
period $q_s q_0/|q_s p_0-p_s q_0|$. The $k_2$ in (\ref{eq:offdiag}) becomes
$k_{2,n}=k_2+2\pi n((p_s q_0-p_0 q_s)/q_0 q_s)$ and $k_1$ is also
site-dependent.
This approach generalizes to $t_{ab}\neq 0$ with the result that singular
energies are the $q_0$ roots of the equation
\be   P_0(E)-4(-1)^{p_0}t_{ab}^{q_0}=0\label{eq:singular1}\ee
for $t_a=t_b$ dominant and of
\be   P_0(E)-4(-1)^{p_0}t_{ab}^{q_0}
      T_{q_0}({t_a\over 2t_{ab}})T_{q_0}({t_b\over 2t_{ab}})=0
\label{eq:singular2}\ee
for $t_{ab}=t_{a\bar b}$ dominant regime.
We have done some numerical check for $p_0=1,q_0=2$ and found that the
singular energies come at $\pm 2t_a$ and
$\pm 2\sqrt{t_a^2+t_b^2-t_a^2t_b^2}$ respectively in agreement with
predictions given by Eqs.(\ref{eq:singular1}) and (\ref{eq:singular2}).

\section{Discussion}

Our studies have shown that the characteristic critical properties of
Harper's equation persist when the symmetry is broken by terms which couple
sites to their next nearest neighbors, provided that the reflection symmetry
in the diagonals is preserved, and provided that the next nearest neighbor
terms satisfy the inequality $t_{ab}+t_{a\bar b}<t_a=t_b$. In this region
the multifractal analysis gives a universal result, strict bounds
for the width $W$ of the spectrum show that it must go to zero with the
reciprocal of the denominator $q$, and numerical results and semiclassical
analysis show that $qW$ has a limit which is rescaled by the next nearest
neighbor coupling, but which is independent of the numerator $p$. It is only
in the corrections to scaling that any oscillatory behavior shows up.

For the bicritical line $t_{ab}=t_{a\bar b}={\rm max}(t_a,t_b)/2$, which
divides the region dominated by the nearest neighbor terms from that
dominated by the next nearest neighbor terms, the multifractal behavior is
quite different, and the width $W$ of the spectrum is known to be
proportional to $1/q^2$, as we know from the rigorous
bounds,\cite{last92,last93} as well as from numerical work. In this case an
oscillatory behavior superposed on the $1/q^2$ dependence shows up, whose
periodicity is at least partially understood from the WKB analysis of
Sec.~VI.

The region for which we have least understanding is the region dominated by
the next nearest neighbor terms, with $t_{ab}=t_{a\bar b}>{\rm
max}(t_a,t_b)/2$. Although the case $t_a=0=t_b$ is equivalent to the
original Harper equation, with the axes turned by $\pi/4$ and the unit cell
doubled in area, for any nonzero values of the nearest neighbor coupling
their strengths $t_a,t_b$ remain relevant, as can be seen clearly from
Eq.~\ref{eq:offdiag}. The spectrum has a multifractal form, but the
multifractal analysis gives different ranges of the exponent for different
values of the parameters of the Hamiltonian. There seems to be a marked
difference between the behavior for $t_a=t_b$ and for $t_a\ne t_b$. When the
width $W$ of the spectrum is studied the results are quite different for the
ratio of Fibonacci numbers and for $1/q$ or similar sequences of fractions
with fixed numerator. For $\phi=1/q$ we find $W$ going to zero like $1/q$,
with an oscillatory coefficient, while for the Fibonacci sequence the
oscillatory behavior is similar, but the dependence on the denominator is of
the form $1/q^\delta$, where $\delta$ seems to be about 1.25 for $t_a=t_b$,
and 1.56 for for $t_a\ne t_b$.

We believe a partial understanding of this behavior can be obtained from our
results for $\phi$ of the form $1/(q_1+1/q_2)$ which were discussed in
Sec.~\ref{bw3}. For large values of the numbers in the continued fraction
expansion
of $\phi$ we get relatively simple scaling behavior which can be studied by
using WKB theory, although we have not carried out the analysis in detail
yet. Because of the periodic or nearly periodic behavior as the $q_j$ are
varied, each stage of the scaling may carry one arbitrarily close to the
bicritical boundary of the critical region, and we found examples, one of
which is shown in Fig.~\ref{fig:han4}, where for certain values of $q_1$ the
dependence of $W$ on $q_2$ at the next stage of scaling was the $1/q_2^2$
typical of the bicritical boundary. Approximants of the golden mean which
are the ratios of two Fibonacci numbers have continued fraction expansions
in which the terms are particularly small, so one is very far from the
simple scaling behavior expected for large $q_j$, so the scaling behavior at
each stage may be intermediate between the critical and the bicritical
behavior.

\acknowledgments
We are particularly grateful to Yoram Last for illuminating discussions,
and for pointing out that the upper and lower bounds obtained in Ref.\
\ref{cit:last} are almost immediately applicable to this problem.

\appendix\section{Evaluation of product of off-diagonal terms}

Evaluation of the product in Eq.~(\ref{eq:offd}) is required to obtain
Eq.~(\ref{eq:offdiag}). The expression
\be
 P(k_2)=\prod_{n=0}^{q-1}\bigl(t_a+t_{a\bar b}e^{-{2\pi ip\over q}
(n+{1\over 2})-ik_2} +t_{ab}e^{{2\pi ip\over q}(n+{1\over 2})+ik_2} \bigr)
\label{eq:app1} \ee
must be periodic in $k_2$ with period $2\pi/q$, since the addition of
$2\pi/q$ to $k_2$ yields a permutation of the same factors in the product.
This tells us that the only terms that survive in the product are those
with $q$ factors of $\exp(-ik_2)$, those with $q$ factors of $\exp(ik_2)$,
and those with equal numbers of factors of $\exp(-ik_2)$ and $\exp(ik_2)$.
The first two cases give
\be
 t_{a\bar b}^q e^{-\pi ipq-iqk_2} + t_{ab}^q e^{\pi ipq+iqk_2}\;,
\label{eq:app2} \ee
while the $k_2$ independent terms can be written as
\be
Q= \prod_{n=0}^{q-1}\bigl(t_a+te^{-{2\pi ip\over q}
(n+{1\over 2})-ik_2} +te^{{2\pi ip\over q}(n+{1\over 2})+ik_2} \bigr)
-(-1)^{pq}2t^q\cos(qk_2)\;,
\label{eq:app3} \ee
where $t^2=t_{a\bar b}t_{ab}$; here we have subtracted off the $k_2$
dependent terms using Eq.~(\ref{eq:app2}). The expression obtained from
Eq.~(\ref{eq:app3}) by setting $k_2=0$,
\be
Q/t^q+(-1)^{pq}2= \prod_{n=0}^{q-1}\bigl(t_a/t+e^{-{2\pi ip\over q}
(n+{1\over 2})} +e^{{2\pi ip\over q}(n+{1\over 2})} \bigr) \;,
\ee
is a polynomial in $t_a/t$ whose zeros are given by
\be
{t_a\over t}=-2\cos[{2\pi p\over q}(n+{1\over 2})]\;.
\ee
These are the zeros of the equation
\be
\cos(q\arccos{t_a\over 2t})\equiv T_q({t_a\over 2t})=(-1)^{p-q}=-(-1)^{pq}\;,
\ee
where $T_q$ is the Chebyshev polynomial of order $q$. Since the coefficient
of $x^q$ in $T_q(x)$ is $2^{q-1}$, this gives
\be
Q/t^q+(-1)^{pq}2=2T_q(t_a/2t)+(-1)^{pq}2\;.
\ee
Combination of this with Eq.~(\ref{eq:app2}) in Eq.~(\ref{eq:app1}) gives
\[
 P(k_2)=\prod_{n=0}^{q-1}\bigl(t_a+t_{a\bar b}e^{-{2\pi ip\over q}
(n+{1\over 2})-ik_2} +t_{ab}e^{{2\pi ip\over q}(n+{1\over 2})+ik_2} \bigr)
\]
\be
=(-1)^{p-q-1} [t_{a\bar b}^q e^{-iqk_2} + t_{ab}^q e^{iqk_2}]
+2(t_{a\bar b}t_{ab})^{q/2}T_q\bigl({t_a\over 2\sqrt{t_{a\bar b}t_{ab}}}
\bigr) \;,
\label{eq:app8} \ee
which is the result used to derive Eq.~(\ref{eq:offdiag}).

\section{Exact expressions for the intersection spectrum}

Avron, Mouche and Simon\cite{avron90} have shown that for the case
$t_{a\bar b}=0=t_{ab}$ the intersection spectrum has measure $4|t_b-t_a|$.
By exploiting the techniques used in earlier
work\cite{thouless83,thouless91b} we can generalize this result to the
case $t_b\ge t_a>t_{a\bar b}+t_{ab}$. This argument depends in its details
on the parities of $p$ and $q$, so we will give the argument explicitly
for the case of $p$, $q$ odd. For $qk_2=\pi$ and $qk_1$ zero or $\pi$ we
exploit the symmetry of Eq.~({\ref{eq:nnn}) about the points $n=0$ and
$n=q/2$, which allows the problem to be reduced to eigenvalue problems
for tridiagonal matrices of order $(q\pm 1)/2$. The eigenvalues $E_m^{++}$
and $E_m^{--}$ which correspond to eigenstates of Eqs.~(\ref{eq:nnn}) and
(\ref{eq:ft}) that are even about both these symmetry points or odd about
both of them give solutions with $qk_1=0$, while $E_m^{+-}$ and $E_m^{-+}$
give solutions with $qk_1=\pi$. We number the eigenvalues from highest to
lowest, starting with zero for the solutions $E_0^{++}$ and $E_0^{+-}$
which are even about $n=0$, and starting with unity for those that are odd
about this point. The highest value of $m$ is $(q-1)/2$ in all four cases.
The matrices corresponding to odd and even boundary conditions about the
point $n=0$ differ only in whether the 01 element is zero or not, and the
trace formula gives
\be
2t_b-E_0^{+-}+\sum_m( E_m^{--}-E_m^{+-}) =0\;,
\label{eq:appb1}\ee
where each term in the sum is positive. The matrices corresponding to odd
and even boundary conditions about $n=q/2$ differ in having an extra term
$\pm (t_a-t_{a\bar b}-t_{ab})$ in the lowest diagonal element, so the trace
formula gives
\be
\sum_m( E_m^{-+}-E_m^{--}) =2 (t_a-t_{a\bar b}-t_{ab})\;,
\label{eq:appb2}\ee
with each term in the sum positive. Addition of these two equations gives
\be
\sum_m( E_m^{-+}-E_m^{+-}) =E_0^{+-}-2 (t_b-t_a+t_{a\bar b}+t_{ab})\;.
\label{eq:appb3}\ee
The terms in the sum are all positive, and are equal to the band gaps
with $qk_2=0$, $qk_1=\pi$ in the intersection spectrum, while the
eigenvalue $E_0^{+-}$ is the highest band edge of the intersection
spectrum. A similar argument can be constructed for the solutions with
$qk_2=\pi$, $qk_1=0$, where we number the eigenvalues ${\cal E}_m^{++}$,
 ${\cal E}_m^{--}$ in increasing order. This gives
\be
\sum_m({\cal E}_m^{++}-{\cal E}_m^{--}) =-{\cal E}_0^{++}-2
(t_b-t_a-t_{a\bar b}-t_{ab})\;.
\label{eq:appb4}\ee
Again, all the terms in the sum are positive and give band gaps of the
intersection spectrum, while ${\cal E}_0^{++}$ is the lowest band edge of
this spectrum. Addition of Eqs.~(\ref{eq:appb3}) and (\ref{eq:appb4})
gives the measure of the intersection spectrum as
\be
E_0^{+-}-{\cal E}_0^{++}- \sum_m( E_m^{-+}-E_m^{+-})- \sum_m({\cal
E}_m^{++}-{\cal E}_m^{--}) =4 (t_b-t_a)\;.
\label{eq:appb5}\ee

For the case $t_b\ge t_{a\bar b}+t_{ab}> t_a$ there is no change in the
argument that leads to Eq.~(\ref{eq:appb4}), but
Eqs.~(\ref{eq:appb1})--(\ref{eq:appb3}) must be replaced by
\be
2t_b-E_0^{++}+\sum_m( E_m^{-+}-E_m^{++}) =0\;,
\label{eq:appb6}\ee
\be
\sum_m( E_m^{--}-E_m^{-+}) =2 (t_{a\bar b}+t_{ab}-t_a)\;,
\label{eq:appb7}\ee
and
\be
\sum_m( E_m^{--}-E_m^{++}) =E_0^{++}-2 (t_b+t_a-t_{a\bar
b}-t_{ab})\;,
\label{eq:appb8}\ee
with all terms in the sums positive. Addition of this to
Eq.~(\ref{eq:appb4}) gives the measure of the spectrum for $qk_1=0$ as
\be
E_0^{++}-{\cal E}_0^{++}- \sum_m( E_m^{--}-E_m^{++})- \sum_m({\cal
E}_m^{++}-{\cal E}_m^{--}) =4 (t_b-t_{a\bar b}-t_{ab})\;,
\label{eq:appb9}\ee
and this gives the generalization of the result for the intersection
spectrum for the case $t_b\ge t_{a\bar b}+t_{ab}\ge t_a$.

\begin{figure}
\caption{Phase diagram of the different regions of the parameters of the
Hamiltonian, for $t_{ab}=t_{a\bar b}$. In region I  states are pure
extended, and
in region III the states are purely localized states. On
the line $BC$ the Hamiltonian has both diagonals  as reflection lines, and the
behavior is critical, very similar to that at the Harper point $B$. Bicritical
behavior is found not only at the point $C$, but along the lines $AC$
and $CE$.}
\label{fig:hiramoto1}
\end{figure}

\begin{figure}
\caption{(a) Spectra for various points on line $BD$ in
Fig.~\protect\ref{fig:hiramoto1} ($t_a = t_b$). (b) Enlarged version of (a).}
\label{fig:hiramoto2}
\end{figure}

\begin{figure}
\caption{ (a) Plot of $f(\alpha )$ for $t_{ab} =0.4t_a$, $t_a=t_b$.
Other points on the line $BC$ of Fig.~\protect\ref{fig:hiramoto1} give
identical results. (b) Plot of $f(\alpha)$ for the bicritical point
$C$ where $t_{ab} =0.5t_a$, $t_a=t_b$.}
\label{fig:hiramoto3}
\end{figure}

\begin{figure}
\caption{ Plot of $F_n^2 W$ versus $n$ for $t_{ab}/t_a=0.5$ and
various values of $t_b/t_a$.  A period 4 oscillation as a function of
$n$ for the case $t_b/t_a=0.5$ can be seen clearly superposed on the
$1/F_n^2$ dependence of $W$.}
\label{fig:hiramoto4}
\end{figure}

\begin{figure}
\caption{Plots of  $F_n^{1.25} W$ on a logarithmic scale versus
$n$ for $t_a=t_b$ and various values of $t_{ab}/t_a$. Period 6 can
be seen for $t_{ab}/t_a=1/\protect\sqrt{2}$, and period 4 for $t_{ab}/t_a=1$.}
\label{fig:hiramoto5}
\end{figure}

\begin{figure}
\caption{ Plots of  $F_n^{1.56} W$ on a logarithmic scale versus $n$ for
various
values of $t_{ab}/t_a$ and $t_a/t_b\ne 1$. For $t_{ab}/t_a=1/\protect\sqrt{2}$
a period of 6 can be clearly seen.}
\label{fig:hiramoto6}
\end{figure}

\begin{figure}
\caption{Plots of $qW/\protect\sqrt{1-4t_{ab}^2}$ as a function of $q$ for
$t_{ab}$=0, 0.1 with $t_a=t_b$=1. The squares representing
$t_{ab}$=0 form envelopes (upper and lower bounds) for other
values of $t_{ab}<0.5$. }
\label{fig:han1}
\end{figure}

\begin{figure}
\caption{Plots of $qW/\protect\sqrt{(1-t_a^2)(1-t_b^2)} \; $ as a function of
$q$   in the critical regime
$2t_{ab}=1>t_a,t_b$ for several values of
$(t_a,t_b)=(\cos(\pi p_1/q_1),\cos(\pi p_2/q_2))$. Each graph is
characterized by a set of fractions, $(p_1/q_1,p_2/q_2)$.}
\label{fig:han2}
\end{figure}

\begin{figure}
\caption{Plots of $q^2 W$ as a function of $q$ at the bicritical point
$2t_{ab}$=1=$t_a$ for $t_b$=0.5 and 0.99 respectively.}
\label{fig:han3}
\end{figure}

\begin{figure}
\caption{Plot of $qW$ on a logarithmic scale against $\ln q$ for
$\phi=p/q=p/(p^2+1)$. Points for $p=2\bmod 4$ are clearly seen to lie
on a line with a slope -0.50. }
\label{fig:han4}
\end{figure}

\begin{table}
\caption{Table of the extremal values of $\alpha$ and the position of
the maximum in the multifractal analysis for various values of the
parameters of the Hamiltonian.}
\begin{tabular}{cccccc}
$t_{ab}/t_a$& $t_b/t_a$& $\alpha_{min}$& $\alpha_{max}$& $\alpha_0$&
$f(\alpha_0)$\\
\tableline

0,.2,.4& 1.0& .421& .547& .50& .50\\
.5& 1.0& .272& .421& .33& .32\\
.5& .75& .282& .381& .35& .34\\
.5& .5& .281& .366& .34& .34\\
.5& .25& .281& .370& .33& .32\\
1.0& 1.0& .300& .650& .43& .41\\
2.0& 1.0& .305& .640& .42& .41\\
3.0& 1.0& .313& .628& .44& .42\\
.6& .5& .29& .41& .36& .35\\
\end{tabular}






\end{table}

\end{document}